%% file: nime-template.tex
\documentclass{nime-alternate}

\usepackage{anonymize}

\usepackage[utf8]{inputenc}

\usepackage{float}

\begin{document}
\conferenceinfo{NIME'23,}{31 May–2 June, 2023, Mexico City, Mexico.}

\title{SnakeSynth: New Interactions for Generative Audio Synthesis}
\label{key}

\numberofauthors{1}
\author{
  \alignauthor \anonymize{Eric Easthope} \\
      \affaddr{\anonymize{University of British Columbia}} \\
      \affaddr{\anonymize{Vancouver, British Columbia, Canada}} \\
      \email{\anonymize{mail@eric.cyou}}
}

\maketitle

\begin{abstract}
    \input{./src/abstract.tex}
\end{abstract}




\section{Background}
\input{./src/background.tex}

\section{Design}
\input{./src/design.tex}

\section{Discussion}
\input{./src/discussion.tex}

\section{Conclusion}
\input{./src/conclusion.tex}

\section{Acknowledgments}
\anonymize{
  I thank Robert for his guidance and for pointing to the novelty of real-time GAN synthesis in the browser.
}

\section{Ethical Standards}
The author is self-funded and reports no conflicts of interest. No living subjects were studied in this work.

\bibliographystyle{abbrv}
\bibliography{refs}

\clearpage

\appendix
\section{Interactions}
\label{appendix:interactions}

\begin{figure}[htbp]
    \centering
        \includegraphics[width=1\columnwidth]{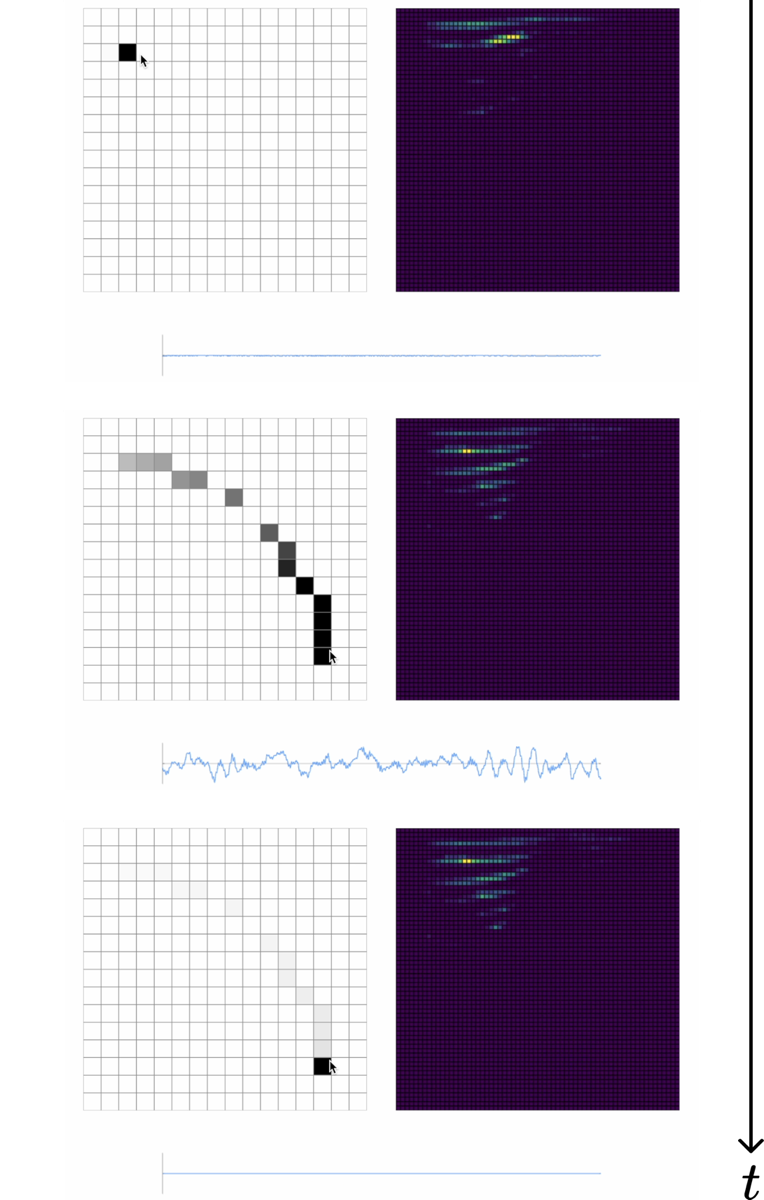}
    \caption{Linear or near-linear gestures produce variable-length audio (resembles ``strumming''). Gesture distance determines sound length.}
    \label{fig:4a}
\end{figure}

\begin{figure}[htbp]
	\centering
		\includegraphics[width=1\columnwidth]{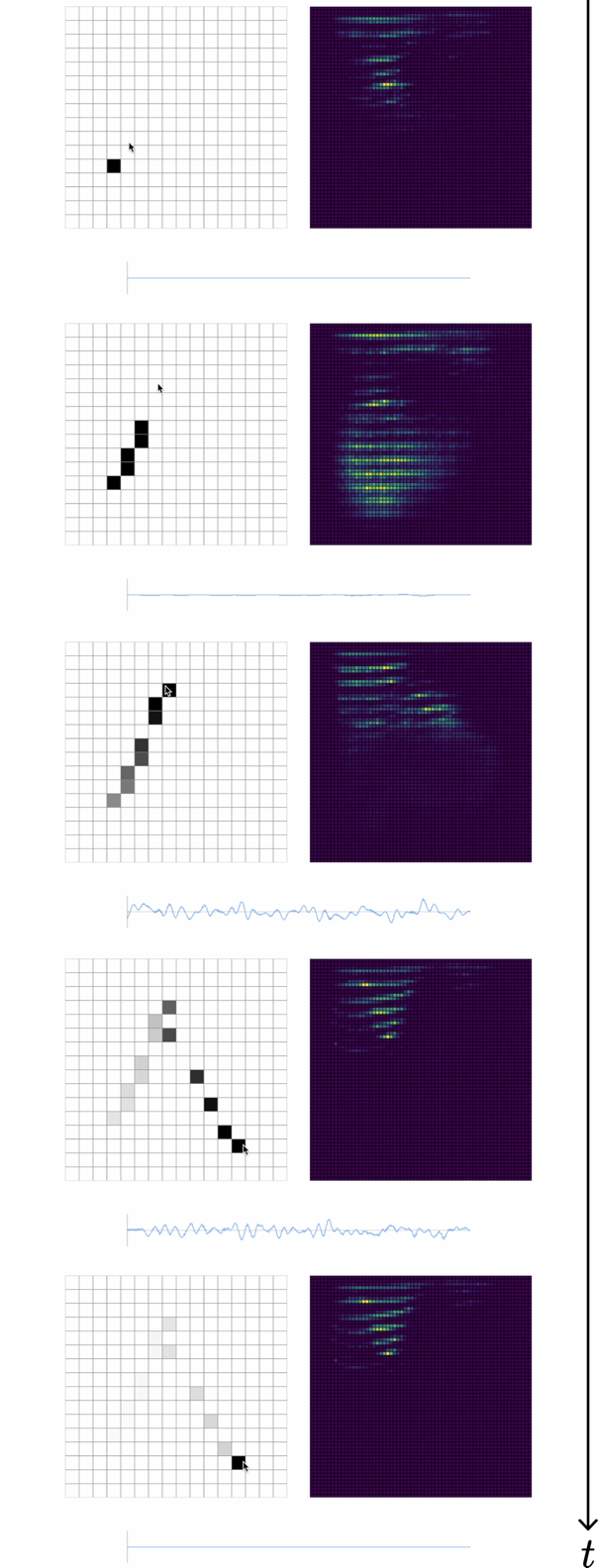}
	\caption{Suddenly changing movement direction creates sudden audio changes and corresponding audio attack (resembles a ``finite bow'').}
	\label{fig:4b}
\end{figure}

\begin{figure}[htbp]
    \centering
        \includegraphics[width=1\columnwidth]{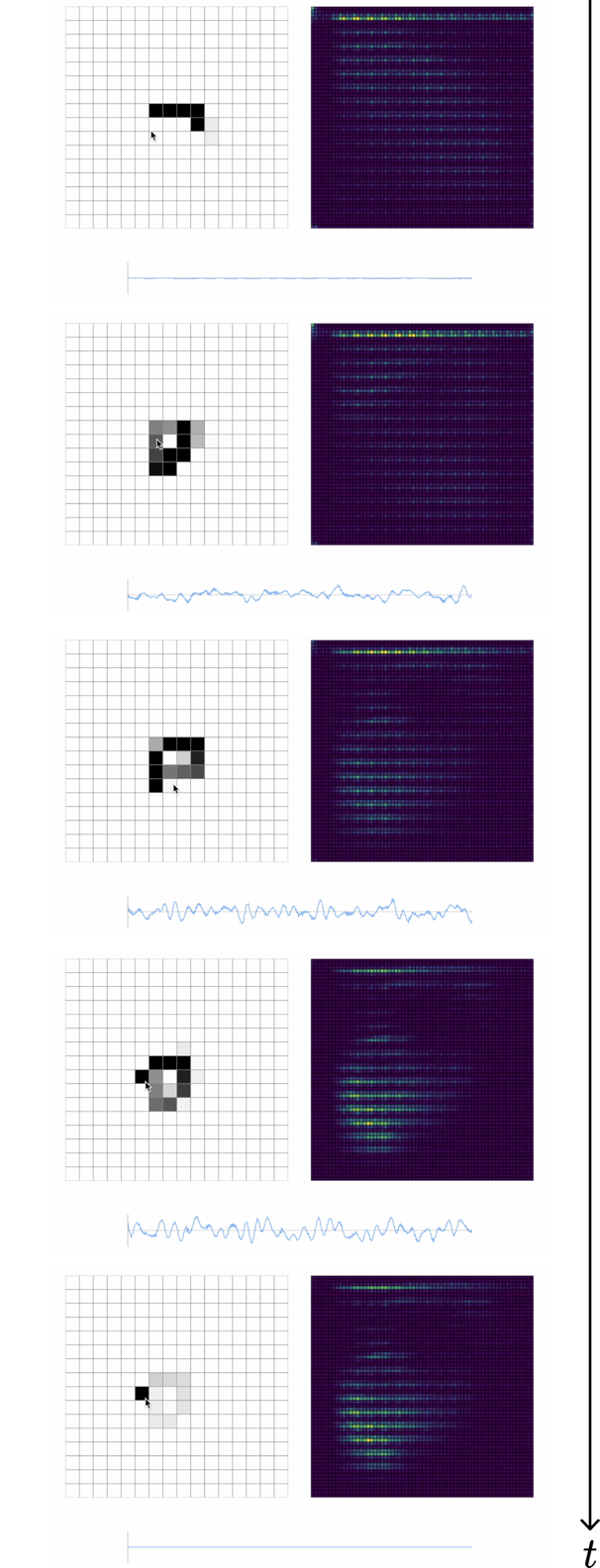}
    \caption{Continuous gestures create continuous audio (resembles an ``infinite bow''). Particularly circular or near-circular gestures produce continuous \emph{rhythmic} audio.}
    \label{fig:4c}
\end{figure}

\begin{figure}[htbp]
	\centering
		\includegraphics[width=1\columnwidth]{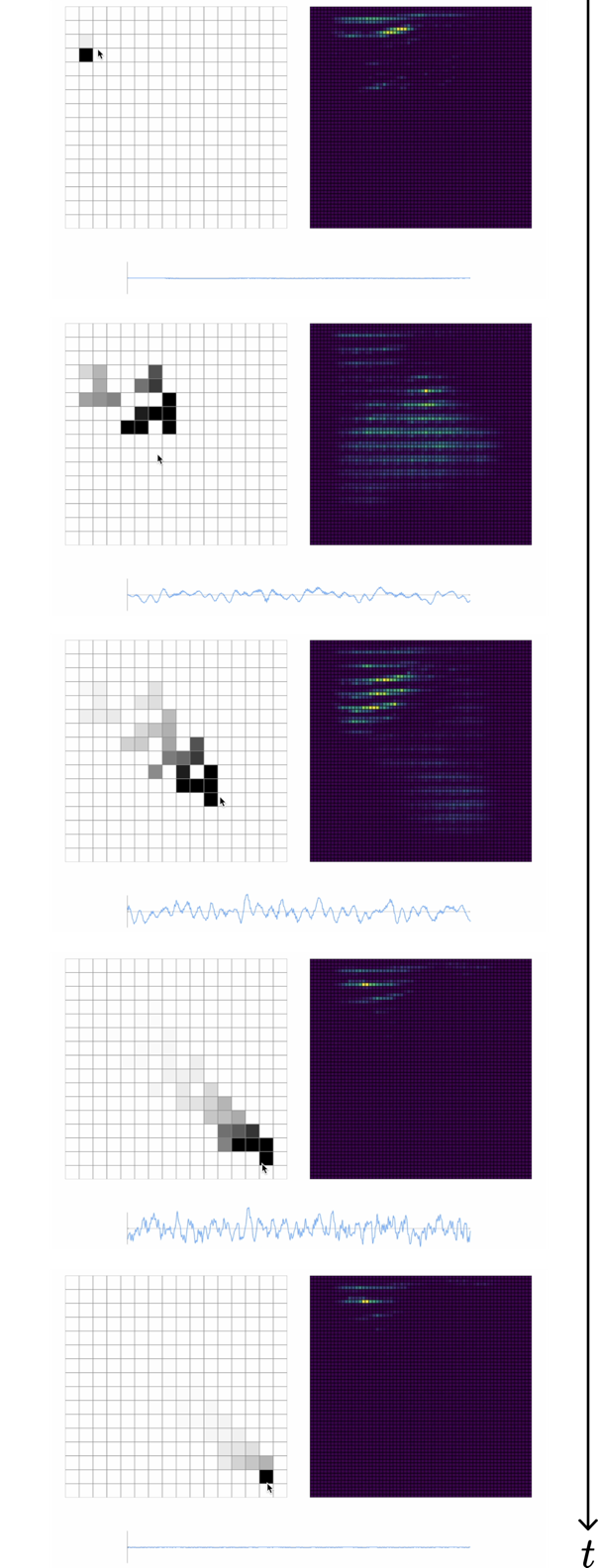}
	\caption{Chaotic gestures with many directional changes to linear and/or circular movements create cacophonous audio (resembles ``brushing'').}
	\label{fig:4d}
\end{figure}


\end{document}

%% file: src/abstract.tex
 I present \emph{SnakeSynth}, a web-based lightweight audio synthesizer that combines audio generated by a deep generative model and real-time continuous two-dimensional (2D) input to create and control variable-length generative sounds through 2D interaction gestures. Interaction gestures are touch and mobile-compatible with analogies to strummed, bowed, and plucked musical instrument controls. Point-and-click and drag-and-drop gestures directly control audio playback length and I show that sound length and intensity are modulated by interactions with a programmable 2D coordinate grid. Leveraging the speed and ubiquity of browser-based audio and hardware acceleration in Google's \texttt{TensorFlow.js} we generate time-varying high-fidelity sounds with real-time interactivity. SnakeSynth adaptively reproduces and interpolates between sounds encountered during model training, notably without long training times, and I briefly discuss possible futures for deep generative models as an interactive paradigm for musical expression.

%% file: src/background.tex
Interaction paradigms for deep generative models (DGMs) have remained relatively shallow in contrast to the diversity of interactions that are possible with musical interfaces and most research interest in DGMs still seems to revolve around generation of fixed-size images and audio in correspondence to fixed-size training data \cite{GANArtSurvey, GANMusicSysReview}. These models often work by learning a low-dimensional set of inputs that resemble the statistics of a training dataset enabling us to generate new samples through a small number of controls significantly smaller than the size of training data. In experimental contexts DGMs seem to be capable of generating novel outputs and controllably interpolating between training data features \cite{GANspace}. Recent developments including the success of WaveNet \cite{WaveNet} and GANsynth \cite{GANsynth} have revealed possibilities for how DGMs might be developed to be more expressive in terms of their outputs and this has consolidated some interest in using DGMs as tools for musical expression. Possibly the largest unified effort to do this might be Magenta (\url{https://research.google/teams/brain/magenta/}) at Google Research which leverages DGMs as part of a larger effort to create music synthetically using machine learning (ML) models.

Yet many projects featured for DGM-based music production and performance still suffer from common structural limitations in DGMs and how they function. Auto-regressive models \cite{AR} like WaveNet \cite{WaveNet} inherently rely on sequential and often somewhat random updates to inputs to produce appreciable changes in outputs. In performance contexts changes in sound in response to new inputs then need to computed in real time or otherwise delayed. Setting aside the challenges of computing DGM outputs in real-time this breaks down an essential auditory feedback loop between a performer and their instrument(s). Responses on the part of the performer in response to an auto-regressive DGM then have to be anticipated as inputs sequentially and must randomly evolve towards more refined outputs.

This runs counter to how we think about musical instruments and related interfaces. While we would not expect a string to resonate the same way every time the same note is played, we do expect to hear the same note and for it to resonate when we play it. Correspondingly there is some expectation in performance being informed by anticipation about where and how to sound instruments in a one-to-one way. This one-to-one-ness also ensures that instruments play the same way today as they do tomorrow. In turn we think that regressive DGMs and randomness alone cannot produce usable and moreover re-usable ML-based digital music tools.

Luckily not all deep generative models are regressive and some are capable of producing inputs and outputs in a one-to-one way. Generative Adversarial Networks (GANs) \cite{GAN} and variational auto-encoders \cite{VAE} amongst other DGMs require only a single forward pass (``single-pass'') from input to output making them better candidates for musical interfaces by enabling performers to learn relationships between how to play digital instruments and what will be sounded when they play them. The development of audio-based GANs by Donahue et al.~\cite{WaveGAN} and Engel et al.~\cite{GANsynth} have shown particular promise for generating novel sounds and musical forms. Technically speaking \texttt{(\text{input}, \text{output})} pairs can be established and anticipated during performance.

In broader performance contexts DGM-based instruments should
exhibit compatible playing dynamics with respect to player expectation.
The application of more energy to the instrument, for example by
``strumming'' or ``bowing'' with greater intensity, should correspondingly
produce more albeit possibly cacophonous sound. Moving or scanning to selectively ``pluck'' strings should not produce unwanted sound. Continuous ``bowed'' sounds should correspond to continuous movements, particularly mechanical ``driving'' and resonance. Reversing
the direction of movement should ``reverse'' the sound in some way; on a
string this might correspond to differences in ``down-picked'' and ``up-picked''
sounds. These are difficult to express with DGMs and even neural
networks broadly speaking due to the fixed length of their
outputs and so there is an opportunity here for new designs. The key problem
is making DGMs expressive in ways beyond their capacity to yield
different outputs.

Part of this is a matter of producing and controlling continuous
variable-length sounds with DGMs. Discrete trigger-based controls for
musical DGMs resembling MIDI inputs are common but offer little to no
control over the length of output audio. This puts the burden of
controlling audio length on the underlying DGM(s). Previous work has
done little to address the generation of variable-length audio with DGMs
despite the apparent utility of producing variable-length sounds in
music contexts. We can concatenate sounds to produce longer streams of
audio but results are often cacophonous (see algorithmic music from
Dadabots, \url{https://dadabots.com}). Thinking in terms of a 2D image
DGM architecture this is equivalent to generating variable-width images
by conjoining images sequentially and is a patch for the problem at
best. This makes the problem of generating variable-length audio with
DGMs an interesting gap in current work and a means to explore DGM
expressivity in creative settings as a performance tool.

\begin{figure*}[ht]
	\centering
		\includegraphics[width=1\textwidth]{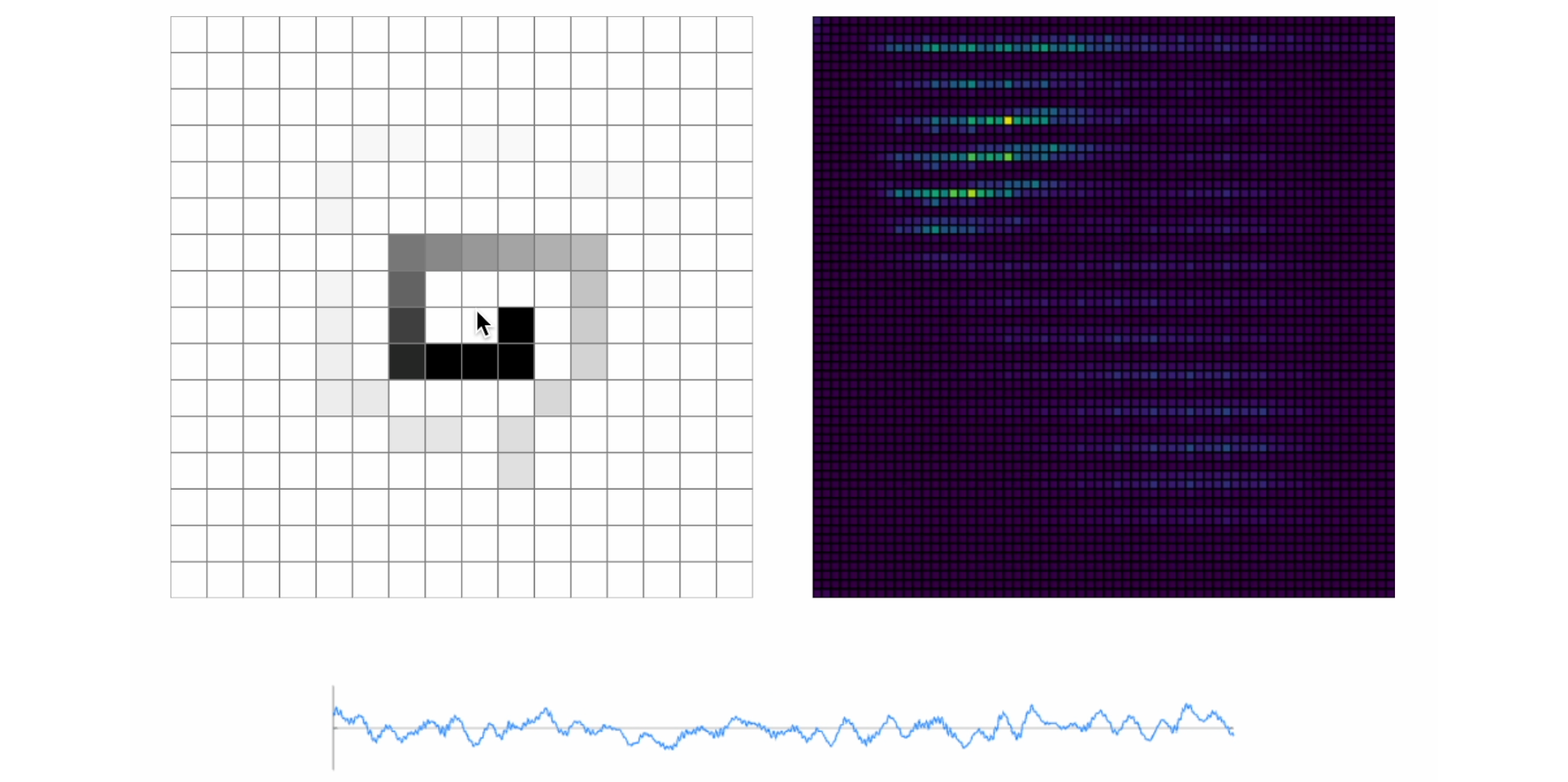}
    	\caption{The cursor-based \emph{SnakeSynth} web interface. Interactions with the grid (left) are also touch-compatible supporting tap-to-click and touch-and-drag gestures in correspondence to point-and-click and drag-and-drop cursor gestures. Mel spectra (right) and time series (bottom) for individual sound samples are displayed and updated in real time during playback.}
    	\label{fig:1}
\end{figure*}

\emph{SnakeSynth} (Figure \ref{fig:1}) is an ML-driven music performance tool and interactive
controller that bridges the gap from discrete trigger-based DGM controls
to continuous variable-length controls to enable new forms of musical
expression and performance dynamics with DGMs. Deriving its name and
interactive paradigm from the ``Snake'' video game genre, SnakeSynth
uses real-time 2D point-and-click and drag-and-drop gestures to directly
control DGM audio playback length to generate variable-length audio in
creative contexts. Interactions with a programmable sound ``grid''
determine audio length relieving DGMs of un-needed extraneous design
constraints and giving more control to performers as the primary
creative agent. We forego concatenation-based approaches and model the
variability of audio length as an external interactive control over what
is otherwise a fixed-length DGM. In turn SnakeSynth enables ``plucked,'' ``strummed,'' and ``bowed'' playing gestures by triggering different fixed-length DGM-generated sounds and blending them through interaction to form longer variable-length sounds.

%% file: src/design.tex
\subsection{Model}
\subsubsection{Generative Adversarial Network}

We set up a GAN made of two networks, a \emph{generator} and a \emph{discriminator}, configured as adversaries such that the generator network learns to generate ``fake'' but convincingly real outputs that ``fool'' classifications by the discriminator. As they train on new samples the generator improves its weights by back-propagation to produce more ``realistic'' outputs that resemble the statistics of training data. In turn the discriminator improves its weights to discern fake samples from dataset samples. Any losses of the generator are theoretically gains of the discriminator and vice versa, and both networks improve with training.

We use a modified Deep Convolutional GAN (DCGAN) architecture and simplify the DCGAN generator \cite{DCGAN} to three convolutional layers and remove the batch normalization layer succeeding the fully-connected layer. DCGANs have the advantage of using local convolutional layers in place of exclusively fully-connected layers \cite{GAN} significantly reducing the total number of trainable parameters and in turn reducing total training time. The generator consists of a fully-connected layer and a sequence of three filter layers each containing a convolutional layer, a batch normalization layer, and activated with a leaky rectified linear unit (leaky ReLU) layer. Batch normalization layers regularize training samples to increase training stability \cite{DCGAN, BatchNorm}. The final convolutional layer is activated with a \(\tanh\) function layer.

\begin{figure*}[htbp]
	\centering
		\includegraphics[width=1\textwidth]{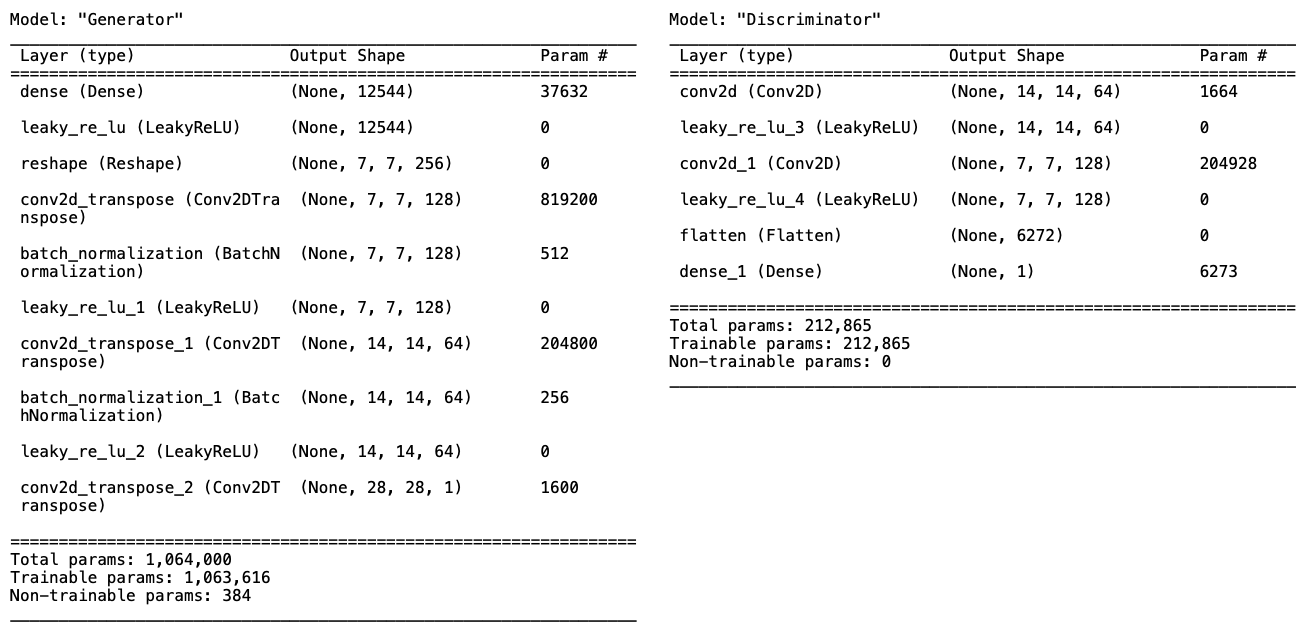}
    	\caption{Simplified Deep Convolutional GAN (DCGAN) \cite{DCGAN} network architecture used to generate SnakeSynth sounds. The generator inputs two values, and outputs 64x64 pixel images activated with a $\tanh$ function layer. The discriminator inputs these 64x64 pixel images, and outputs one value to classify ``real'' dataset samples versus ``fake'' generator samples. The discriminator foregoes an activation function to output values outside of $[-1,1]$.}
    	\label{fig:2}
\end{figure*}

We also use a Convolutional Neural Network (CNN)-based discriminator with two convolutional layers, no dropout layers, and no batch normalization. Again this significantly reduces the total number of trainable parameters. The discriminator consists of two convolutional layers activated with leaky ReLU layers and a fully-connected layer with no activation layer. Without an activation layer the discriminator outputs values outside of \([-1,1]\) and we compute cross-entropy loss directly from \emph{logits} from the fully-connected layer. Unlike the DCGAN authors we do not change any network initialization weights before training and we use \emph{reshape} and \emph{flatten} layers to transform square images to and from layers expecting flat inputs. Both networks are summarized in Figure \ref{fig:2} for a latent space of size two in correspondence with a 2D cursor or touch-based control space. This amounts to a generator with just over one million trainable parameters (approximately 245 parameters per pixel).

\subsubsection{Dataset: 2D Spectral Images}

Our DCGAN training data comprises of square 64x64 pixel images made from Mel-scaled spectral coefficients for a small collection of human voice samples, which is only one of many possible training sets. This is in accordance with observations by Engel et al.~\cite{GANsynth} that GANs are capable of producing high-fidelity audio from limited spectral information and are significantly faster to train. Notably Mel-scaled coefficients are an effective compression of spectral audio information in 2D that simultaneously accounts for human audio perception. Compared to 1D time series-based approaches like WaveGAN \cite{WaveGAN} using images of spectral coefficients in place of audio significantly reduces the overall dimensionality and memory requirements for training and our use of GAN models in SnakeSynth.

Sounds in SnakeSynth are computed by inverting generated 2D spectral images into 1D time series through Griffin-Lim inversion. Each sound is windowed using a cosine window or similar to remove edge audio artifacts. We currently choose the number of sounds in the SnakeSynth interaction grid so this inversion is automated as part of data post-processing. Faster inversions could be realized in real-time interactive settings.

\subsubsection{GAN Training}

Generator outputs are initialized as Gaussian noise and we train the DCGAN generator and discriminator in lockstep, training the generator first to produce new outputs and then the discriminator second on outputs produced by the generator and against real samples from the training dataset. This training strategy is equivalent to a zero-sum competitive game between two players (or networks in this case) where losses of the generator amount to gains of the discriminator and vice versa. Goodfellow et al.~\cite{GAN} represents this with the objective function

\[L(G,D) = \mathbb{E}_{x\sim \mu_{ref}}[\ln D(x)] + \mathbb{E}_{z\sim \mu_{Z}}[\ln(1-D(G(z))]\]

for a generator \(G\) and discriminator \(D\) where \(\mathbb{E}\) shows expectation values and \(\mu_{ref}\) is the set of possible outcomes in the sample distribution and \(\mu_{Z}\) is the set of possible generated outcomes in a Gaussian (normal) distribution. Correspondingly \(\mathbb{E}_{x\sim \mu_{ref}}\) shows the expectation value that real samples (\(x\)) from the training dataset or generated samples (\(z\)) are part of the sample space or the generated space respectively. Together these incur a generator and discriminator ``loss'' during training that we use to back-propagate updates to generator and discriminator weights.

The training dataset is shuffled and separated into batch sizes of one so that every image is seen during training and we train for 300 epochs using the same objective function defined by Goodfellow et al.~\cite{GAN}. Notably increasing the batch size can reduce training time. We find that a small 64x64 pixel DCGAN model with a two-dimensional generator latent space can train hundreds of epochs within a few minutes on a standard MacBook Pro (2020). Re-training on new samples correspondingly takes at most a few minutes longer and trained models and sounds can be stored and loaded for later use without further training.

Because the generator input dimension is only two we are able to directly access and visualize the space of possible generator outputs by passing 2D quantile values as coordinate inputs to the generator latent space. This enables us to produce samples from most of the generator output distribution using a 2D and particularly \emph{finite} grid-based interactive controller. To do this quantile values up to 95th percentile outcomes are computed from the inverse cumulative density function (inverse CDF) for a 2D Gaussian (normal) distribution with zero mean (\(\mu = 0\)) and unit variance (\(\sigma^2 = 1\)). Plotting images generated from these quantile values produces a visualization of nearly all possible generator outputs to arbitrary precision from a restriction of the entire generator output distribution (Figure \ref{fig:3}, right) helping us design controllers with consideration for underlying generator statistics.

\subsection{Interactions}

\begin{figure*}[htbp]
	\centering
		\includegraphics[width=1\textwidth]{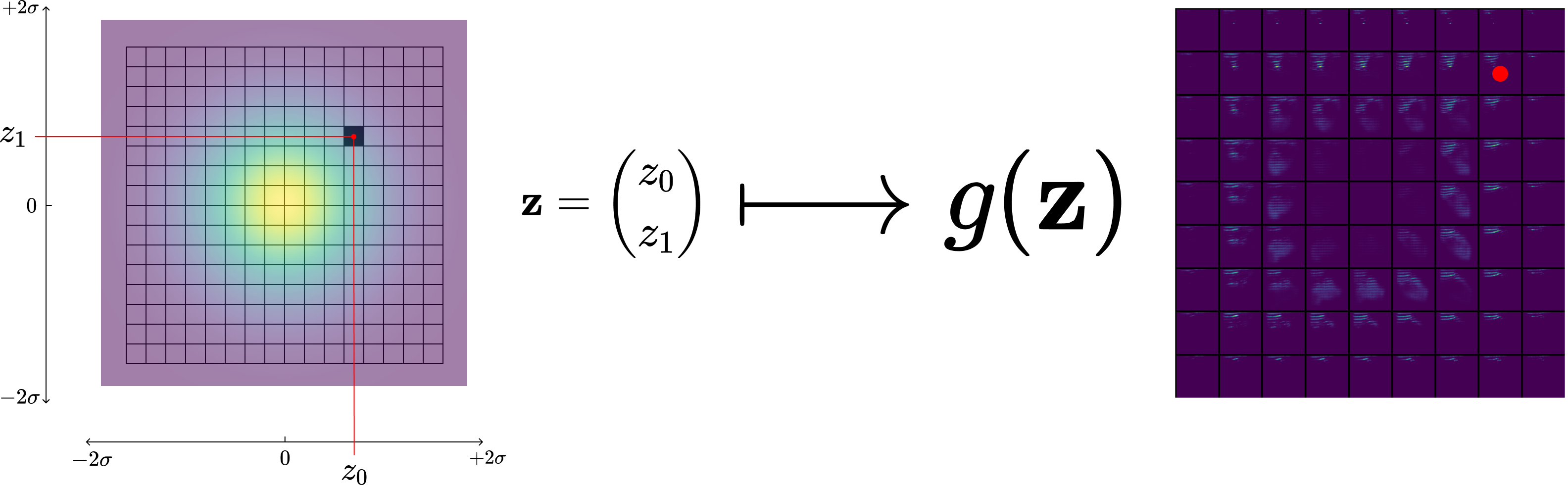}
	\caption{SnakeSynth generates images through maps of Gaussian (normal) quantile values sampled (in red) from two standard deviations in a 2D coordinate grid to a 2D generator latent space. This produces sample outputs from a restriction of the entire possible generator output distribution up to 95th percentile outcomes for different arbitrary grid sizes.}
	\label{fig:3}
\end{figure*}

SnakeSynth affords a number of different interaction types naturally
through interactions with a two-dimensional \(N \times N\) coordinate grid (Figure \ref{fig:3}, left):

\begin{enumerate}
  \item Click (or touch) gestures produce fixed-length audio (resembles ``plucking'').
  \item
    Linear or near-linear gestures produce variable-length audio (resembles ``strumming'') (Figure \ref{fig:4a}). Gesture distance determines sound length.
  \item
    Suddenly changing movement direction creates sudden audio changes and corresponding audio attack (resembles a ``finite bow'') (Figure \ref{fig:4b}).
  \item
    Continuous gestures create continuous audio (resembles an ``infinite bow'')
    (Figure \ref{fig:4c}). Particularly circular or near-circular gestures produce continuous \emph{rhythmic} audio.
  \item
    Chaotic gestures with many directional changes to linear and/or circular movements create cacophonous audio (resembles ``brushing'') (Figure \ref{fig:4d}).
\end{enumerate}

Interactions 2-5 are showcased in Appendix \ref{appendix:interactions}.

\subsection{Synthesis}

Instead of directly concatenating audio clips we trigger equal-length clips asynchronously and sum them over time to produce variable-length audio in response to interaction. Each sound is windowed in data post-processing so this is functionally similar to the overlap-add method. Simulating three equally-spaced interactions with the SnakeSynth grid in Figure \ref{fig:3} we see three generated sample sounds and their windowing functions sum to produce a single variable-length sound. We also see the amplitude of the resulting sound increase with greater sample overlap. This is chosen to produce interaction analogies to mechanical ``driving'' and resonance (as mentioned before) and other ways to blend overlapping audio could be explored.

\begin{figure*}[htb]
	\centering
		\includegraphics[width=1\textwidth]{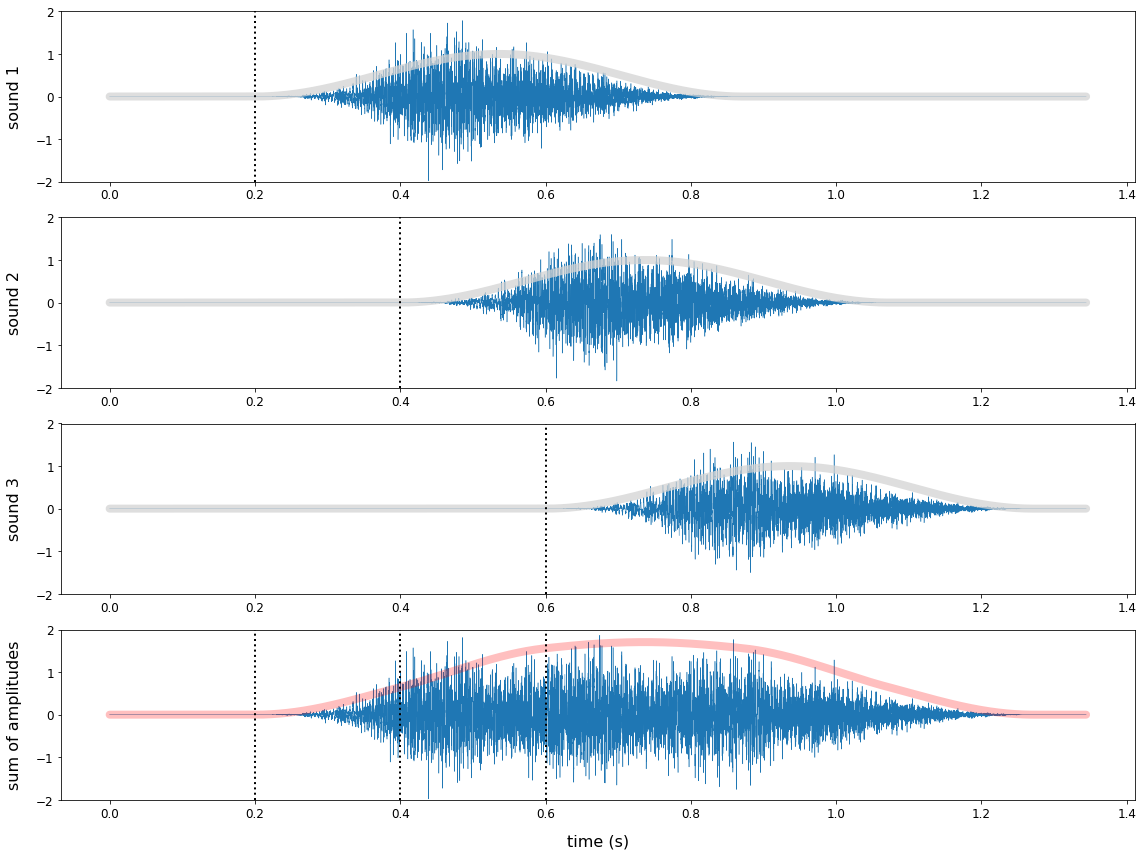}
    	\caption{Three time-triggered sound samples (rows) generated and summed over time to produce a single variable-length sound (bottom). Increasing amplitude by summing overlapping sounds enables virtual mechanical ``driving'' through the SnakeSynth controller to produce resonance-like effects. The amplitude of the resulting sound increases with greater sample overlap.}
    	\label{fig:5}
\end{figure*}

%% file: src/discussion.tex
By foregoing concatenation-based approaches and modelling the variability of audio length in terms of interaction we lose the precision of a triggered fixed-size model and we have to choose how to blend sounds in context. However, we think that what we gain in flexibility in terms of modularity and greater choice over DGMs should not be understated as it keeps the abundance of fixed-length GAN models and ongoing research available to us as design options. Non-generative models of the same dimension would even suffice. Going further this flexibility enables us to create novel controllers for audio DGMs capable of generating variable-length audio. This does not seem to be widely recognized as a design interest and surprisingly we have seen little discussion about it in previous work.

SnakeSynth offers one way around the problem by treating audio length as a parameter of interactive control outside of the generative model. This bridges the gap from fixed-length audio DGMs to controller-driven variable-length DGMs and even to DGM-based music performance by recognizing that asynchronously triggering audio clips over time is congruent to mapping user interactions over time. Given the ubiquity of cursor movements and touch in 2D digital coordinate spaces these seem to be an appropriate starting point for discussion on and exploration of user interaction as a means of DGM control and particularly DGM control for musical expression.

Choices on how to map from the SnakeSynth coordinate grid to generator latent space(s) raise interesting questions about both the shape of the DGM latent spaces themselves but also how to construct novel and/or non-trivial maps between them and the SnakeSynth grid. We are not required to use quantile values as inputs either and interactions could be readily extended to any interface that produces at least two values in real time.

This is somewhere in the design of latent space-based control that existing human-computer interface principles like Fitts' law \cite{FittsLaw} could be applied such as knowledge about distance to target, target size, cognitive load, etc. Similarly the design of real-time sound blending beyond slowed attacks and windowing, especially for asynchronously triggered sounds, deserves further consideration. Semantically some of these design choices would reflect different views of the audio space or context at hand so as to be recognizable and learn-able by performers and reproducible in performance settings.

DGM research continues to evolve at quick pace and we are still finding new ways to train high-fidelity GANs quickly enough, for example by progressively adding layers during training \cite{ProgressiveGAN}, that it may soon be feasible to train small GAN models in real time. This would enable SnakeSynth and derived tools to ``evolve'' new auditory spaces in response to real-time interactions and/or new data. This takes us away from the mindset of fixed-length audio models and re-frames digital musical instruments as things capable of evolving over time to adapt to context and changing how we might think about digital music tools for performance.

%% file: src/conclusion.tex
I showed how \emph{SnakeSynth}, demoed as a web-based audio synthesizer, combines DGM audio and real-time continuous 2D input to create and control variable-length generative sounds through various interaction gestures. Interaction gestures have analogies to strummed, bowed, and plucked musical instrument controls. I showed that sound length and intensity are modulated by interactive control with a 2D programmable sound grid, and briefly discussed possible futures for DGMs as an interactive paradigm for musical expression.

%% file: nime-template.bbl
\begin{thebibliography}{10}

\bibitem{WaveGAN}
C.~Donahue, J.~McAuley, and M.~Puckette.
\newblock Adversarial {Audio} {Synthesis}, Feb. 2019.
\newblock arXiv:1802.04208 [cs].

\bibitem{GANsynth}
J.~Engel, K.~K. Agrawal, S.~Chen, I.~Gulrajani, C.~Donahue, and A.~Roberts.
\newblock {GANSynth}: {Adversarial} {Neural} {Audio} {Synthesis}, Apr. 2019.
\newblock arXiv:1902.08710 [cs, eess, stat].

\bibitem{FittsLaw}
P.~M. Fitts.
\newblock The information capacity of the human motor system in controlling the
  amplitude of movement.
\newblock {\em Journal of Experimental Psychology}, 47(6):381--391, June 1954.

\bibitem{GAN}
I.~J. Goodfellow, J.~Pouget-Abadie, M.~Mirza, B.~Xu, D.~Warde-Farley, S.~Ozair,
  A.~Courville, and Y.~Bengio.
\newblock Generative {Adversarial} {Networks}, June 2014.
\newblock arXiv:1406.2661 [cs, stat].

\bibitem{AR}
K.~Gregor, I.~Danihelka, A.~Mnih, C.~Blundell, and D.~Wierstra.
\newblock Deep {AutoRegressive} {Networks}, May 2014.
\newblock arXiv:1310.8499 [cs, stat].

\bibitem{GANspace}
E.~Härkönen, A.~Hertzmann, J.~Lehtinen, and S.~Paris.
\newblock {GANSpace}: {Discovering} {Interpretable} {GAN} {Controls}, Dec.
  2020.
\newblock arXiv:2004.02546 [cs].

\bibitem{BatchNorm}
S.~Ioffe and C.~Szegedy.
\newblock Batch {Normalization}: {Accelerating} {Deep} {Network} {Training} by
  {Reducing} {Internal} {Covariate} {Shift}.
\newblock In {\em Proceedings of the 32nd {International} {Conference} on
  {Machine} {Learning}}, pages 448--456. PMLR, June 2015.

\bibitem{ProgressiveGAN}
T.~Karras, T.~Aila, S.~Laine, and J.~Lehtinen.
\newblock Progressive {Growing} of {GANs} for {Improved} {Quality},
  {Stability}, and {Variation}, Feb. 2018.
\newblock arXiv:1710.10196 [cs, stat].

\bibitem{VAE}
D.~P. Kingma and M.~Welling.
\newblock Auto-{Encoding} {Variational} {Bayes}, May 2014.
\newblock arXiv:1312.6114 [cs, stat].

\bibitem{WaveNet}
A.~v.~d. Oord, S.~Dieleman, H.~Zen, K.~Simonyan, O.~Vinyals, A.~Graves,
  N.~Kalchbrenner, A.~Senior, and K.~Kavukcuoglu.
\newblock {WaveNet}: {A} {Generative} {Model} for {Raw} {Audio}, Sept. 2016.
\newblock arXiv:1609.03499 [cs].

\bibitem{DCGAN}
A.~Radford, L.~Metz, and S.~Chintala.
\newblock Unsupervised {Representation} {Learning} with {Deep} {Convolutional}
  {Generative} {Adversarial} {Networks}, Jan. 2016.
\newblock arXiv:1511.06434 [cs].

\bibitem{GANArtSurvey}
S.~Shahriar.
\newblock {GAN} computers generate arts? {A} survey on visual arts, music, and
  literary text generation using generative adversarial network.
\newblock {\em Displays}, 73:102237, July 2022.

\bibitem{GANMusicSysReview}
H.~Zhang, L.~Xie, and K.~Qi.
\newblock Implement {Music} {Generation} with {GAN}: {A} {Systematic} {Review}.
\newblock In {\em 2021 {International} {Conference} on {Computer} {Engineering}
  and {Application} ({ICCEA})}, pages 352--355, Kunming, China, June 2021.
  IEEE.

\end{thebibliography}
